\documentclass[%
 preprint,
 superscriptaddress,
 nofootinbib,
 amsmath,amssymb,
 aps
]{revtex4-1}

\usepackage{graphicx,amssymb,bm,latexsym,color,epsf}

\begin{document}


\newcommand{\nn}{\nonumber\\}
\newcommand{\be}{\begin{equation}}
\newcommand{\ee}{\end{equation}}
\newcommand{\bea}{\begin{eqnarray}}
\newcommand{\eea}{\end{eqnarray}}
\newcommand{\G}{\Gamma}
\newcommand{\tr}{\mathrm{tr}}
\newcommand{\Tr}{\mathrm{Tr}}

\newcommand{\La}{\Lambda}
\newcommand{\reg}{\rho_{k, \Lambda}(s)}
\newcommand{\ZL}{Z_\Lambda(\phi)}
\newcommand{\VL}{V_\Lambda(\phi)}
\newcommand{\DS}{\delta S^{(2)}}
\newcommand{\SZ}{S^{(2)}_0}

%
%

\title{Wilsonian Renormalization Group In The Functional Nonperturbative Approach} 

\author{Gian Paolo Vacca}
\affiliation{INFN, Sezione di Bologna, Via Irnerio 46, I-40126 Bologna}

\begin{abstract}
We consider a functional relation between a given Wilsonian RG flow, which has to be related to a specific coarse-graining procedure, and an infinite family of (UV cutoff) scale dependent field redefinitions. 
Within this framework one can define a family of Wilsonian proper-time exact RG equations associated to an arbitrary regulator function. 
New applications of these RG flow schemes to the Ising Universality class in three dimensions in the derivative expansion are shortly illustrated.
\end{abstract}

\maketitle


\section{Introduction}	

Quantum and statistical field theories are at the core of extremely successful descriptions of Physical systems, 
from critical phenomena and phase transition of known microscopical systems which occur effectively at large distances 
to the challenging problem of defining fundamental and consistent theories valid in principle up to arbitrary short distances. 
The modern ideas of renormalization of Field Theories~\cite{Wilson_1975} are providing a way to interpret and classify the space of possible theories, 
where the critical points play a special role. This idea is based on the two step process of coarse-graining and rescaling
which accounts for the process of "integrating out" some of the quantum/statistical fluctuations of the degrees of freedom, 
tipically according to an ordering of their wavelengths (from short to long ones).
The study of renormalization group flows in theory space and of the properties around its fixed points enable us to extract the universal behavior
which can be in common to very different systems. This is in general a challenging task and at analytical level the best tools are given by the so called
exact functional renormalization group techniques\footnote{ 
Purely critical informations can sometime be extracted with conformal field theory techniques if one expects this kind of symmetry enhancement.}.
The Wilsonian action $S_\La$  depends on a UV cutoff $\La$ in such a way that when the former is inserted in a functional integral with a cutoff $\La$, 
the partition function is independent of the cutoff itself, which is just a rephrasing of the fact that physical phenomena do not depend on our choice
of the UV cutoff inserted in the theoretical model.
The corresponding functional RG flow equation were first studied by Wegner and Houghton~\cite{Wegner:1972ih}, then by Polchinski~\cite{Polchinski:1983gv} and later generalized~\cite{Morris:1998kz, Latorre:2000qc, Rosten:2010vm}. 
Wilsonian flows can be introduced also for other functionals. Indeed the infrared regulated effective (average) action, 
related to the one particle irreducible vertices, has been shown to be a very useful description. It satisfy the Wetterich-Morris equation~\cite{wetterich, Morris:1993qb}.

We shall discuss in the following a relation for the flow of the Wilsonian action and consider the inverse problem
of defining which coarse-graining procedures lead to a given functional flow equation~\cite{BLPV}, f
ocusing on the case of a RG flow based on a proper-time (PT) regularization.

\section{General Wilsonian flows}

A Wilsonian action $S_\La$ at the UV scale $\La$ may be defined, as recalled above,  as the action which, when inserted in the UV regulated functional
integral at the scale $\La$, gives the same partition function $Z$ independently on the choice of $\La$:
\begin{equation}
 Z=\int [d\varphi]^\La e^{-S_\La[\varphi]} \quad , \quad \La \frac{\rm d}{{\rm d} \La} Z =0\,,
 \label{partZ}
\end{equation}
which means that the physics predicted is unchanged.

This property can be obtained if the coarse-graining procedure for the Wilsonian action, is such that 
\begin{equation} \label{eq:generalCG}
 \La \frac{d}{d \La} e^{-S_\La[\varphi]}
 = \int \! d x \, \frac{\delta}{\delta  \varphi(x)} \left( \psi^{\La}_{x}[\varphi] e^{- S_{\La}[\varphi]} \right)
\end{equation}
for some family of functionals $\psi_{x}[\varphi]$, since the the partition function $Z$ defined in Eq.~\eqref{partZ} becomes manifestly independent
from $\Lambda$.
One can write this general flow as
\begin{equation} \label{eq:generalFlow}
 \La \frac{d}{d \La} S_{\La}[\varphi]
 = \int d x \left( \frac{\delta S_{\La}[\varphi]}{\delta \varphi(x)} \psi^{\La}_{x}[\varphi]
 - \frac{\delta \psi^{\La}_{x}[\varphi] }{ \delta \varphi(x)} \right) \, ,
\end{equation}
which can be naturally seen as associated to the infinitesimal field redefinition
$\varphi(x) \to \varphi'(x) = \varphi(x) - \frac{\delta \La}{\La} \psi_{x}^{\La}[\varphi]$.
The field redefinition is associated to a coarse-graining
scheme so that the object $\psi_{x}[\varphi]$ is in general non trivially dependent on it and the Wilsonian action itself as well. 
Indeed, considering a coarse-grained step between the scales 
$\La_{0}$ and  $\La$ encoded in the non local mapping $\varphi(x) = b_{\La}[\varphi_{0}](x)$, where $\varphi_{0}$ and the correspoding $S_{\La_{0}}[\varphi_0]$ are associated to the scale $\La_{0}$,  one can for example derive, see~\cite{Latorre:2000qc,Rosten:2010vm},
the following relation
\begin{equation}
 \psi_{x}^{\La}[\varphi]
 = e^{S_{\La}[\varphi]} \int [d \varphi_{0}] \, \delta( \varphi - b_{\La}[\varphi_{0}] ) \,
 \La \frac{ d b_{\La}[\varphi_{0}](x) }{ d \La } \, e^{- S_{\La_{0}}[\varphi_0]} \, .
\end{equation}
A simple example is given by the case of the Wilsonian RG flow considered by Polchinski \cite{Polchinski:1983gv}.
Indeed, from the knowledge of the flow equation, one can directly guess the form of $\psi_{x}^{\La}[\varphi]$ to be plugged in the
general formula~(\ref{eq:generalFlow}):
\begin{equation}
 \psi_{x}^{\La}[\varphi]
 = \frac{1}{2} \int \! d z \, \dot{\Delta}_{x z} \, \frac{ \delta \Sigma[\varphi] }{ \delta \varphi(z)} \, ,
  \label{polc_red}
\end{equation}
where $\Delta$ is a suitably regulated propagator, decaying fast enough at short distances, with a dot standing for the derivative w.r.t.~$\log \La$ and
$\Sigma[\varphi]$ is given by
\be
 \Sigma_{\La}[\varphi] = - \frac{1}{2} \int \! d x \, \varphi(x) (-\Box_{x}) \varphi(x) + S^{I}_{\La}[\varphi]
 \ee
 where we assume the decomposition
 \be
 S_{\La}[\varphi] = \frac{1}{2} \int \! d x \, \varphi(x) (-\Box_{x}) \varphi(x) + S^{I}_{\La}[\varphi] \, .
\ee
This choice leads, as anticipated, to the Polchinski flow equation
\begin{equation} \label{eq:polchinski_eq}
 \La \frac{d}{d \La} S^{I}_{\La}[\varphi]
 = \frac{1}{2} \int \! d x d y \, \dot{\Delta}_{x y}
 \left[ \frac{\delta S^{I}[\varphi]}{\delta  \varphi(y)} \frac{\delta S^{I}[\varphi]}{\delta  \varphi(x)}
 - \frac{\delta^{2} S^{I}[\varphi]}{\delta  \varphi(y) \delta \varphi(x)} \right] \, .
\end{equation}
Let us stress that the same RG flow equation can be also obtained with other slightly different choices of $\psi_{x}^{\La}[\varphi]$, which belongs to an infinite family obtained adding to the expression in Eq.~\eqref{polc_red}
a term $w_x[\varphi] e^{S_\La[\varphi]}$, such that $ \int \! d x \, \frac{\delta}{\delta  \varphi(x)} w_{x}[\varphi] = 0$. 
They just correspond to different but equivalent implementations of the coarse-graining procedure, generating the same flow.

The Polchinski Wilsonian effective action $S_\La$, which satisfies the above flow equation, receives both 1PI (the second term) and 1PR (the first term) contributions.
Only in the sharp cutoff limit, if the momenta flowing into a vertex of a 1PR term have sum below the UV cutoff $\La$, such contributions are absent~\cite{Morris:1993qb}.
This is also the case for the Wegner-Houghton Wilsonian action.
Moreover the Polchinski action has a simple relation with the regulated generator of the connected Green's functions and with the effective average action~\cite{Morris:1993qb}, 
which is the IR regulated generator of the 1PI vertices.

It is therefore interesting to ask if it is possible in general to derive, given a Wilsonian RG flow in some given scheme, the form of $ \psi_{x}^{\La}[\varphi]$ which leads to it.
Having in mind Eq.~\eqref{eq:generalCG} such a possibility can be investigated by construction, looking for a solution of the following equation
\begin{equation} \label{eq:condpsi2}
 \int \! d x \, \frac{\delta}{\delta \varphi(x)} \left( \psi_{x}[\varphi] e^{- S_{\La}[\varphi]} \right)
 = f[\varphi],
\end{equation}
for a specific scheme dependent functional $f[\varphi]$.
%
%
\subsection{Construction of a general solution}
%
%
From the definition recalled above it is clear that such a construction requires the knowledge of the 
Wilsonian action $S_{\La}[\varphi]$ satisfying the RG flow equation, making this problem hard to solve in practice.
Nevertheless a first important step is to know if at least the solution of this problem can exist.

Let us first notice that the left hand side of the Eq.~\eqref{eq:condpsi2} can be seen as a divergence of a vector field belonging to an infinite dimensional vector
space, i.e., has the structure
\begin{equation} \label{infdiv}
 \int \! d x \, \frac{\delta}{\delta \varphi(x)} u_{x}[\varphi] = f[\varphi] \, .
\end{equation}
Keeping in mind that the quadratic kinetic part of the Wilsonian action is defined by a UV regulated Laplacian
$(\Box)_\Lambda$, it is convenient to define a slightly different object $v_{y}[\varphi]$, since, as we shall see, it simplifies the analysis
for the free theory case~\footnote{%
We thank Tim Morris for this suggestion.
}, such that
\begin{equation}
 u_{x}[\varphi] = G_{x y} v_{y}[\varphi] \,,
\end{equation}
where we have introduced the space-time 'regulated' Green's function $G_{x y}$ satisfying
\begin{equation}
 (-\Box_{x})_{\La} G_{x y} = \delta(y) \, .
\end{equation}

One can assume that as in any finite dimensional vector space, the vector field $v_{x}[\varphi]$ can be decomposed as a sum
of a "gradient part" and a divergenceless part as follows
\begin{equation} \label{decomp2}
 v_{x}[\varphi]
 = \frac{\delta}{\delta \varphi(x)} h[\varphi] + (-\Box_{x})_{\La} w_{x}[\varphi], \quad
 \int \! d x \, \frac{\delta}{\delta  \varphi(x)} w_{x}[\varphi] = 0 \, .
\end{equation}

Then a particular solution of Eq.~(\ref{infdiv}) using Eq.~(\ref{decomp2}) can be obtained solving an infinite dimensional
Poisson-like equation
\begin{equation} \label{infpoisson2}
 \int \! d x d y \, \frac{\delta}{\delta \varphi(x)} G_{x y} \frac{\delta}{\delta \varphi(y)} h[\varphi] = f[\varphi] \, .
\end{equation}
where the metric in field space is field independent (flat geometry in field space) even if has a non trivial space-time
dependence.
Since the determinant of the metric is field independent one can see that the operator in field space above is really a
covariant Laplacian.
We expect that this generalized "elliptic" second order linear differential problem can have solutions,
unique at least from the physical point of view when suitable boundary conditions are imposed. 

Let us write a functional "Fourier" transform for the scalar $h$ introducing a source $J$
\begin{equation}
 \tilde h[J] = \int [d \varphi] e^{-i \varphi \cdot J} h[\varphi]
\end{equation}
and similarly for the right hand side of Eq.~(\ref{infpoisson2}), i.e. for $f[\varphi]$.
Then on can formally rewrite this generalized Poisson equation as
\begin{equation} \label{Fourierpoisson1}
 - (J \cdot G \cdot J) \, \tilde h[J] = \tilde f[J] \, ,
\end{equation}
where $J \cdot G \cdot J = \int \! d x d y \, J (x) G_{x y} J(y)$, so that we can write its solution as
\begin{equation}
 h[\varphi] = - \int [d J] e^{i \varphi \cdot J} \frac{1}{J \cdot G \cdot J} \tilde f[J] \, .
\end{equation}

Using this setup for our original problem we can then formally write
\begin{equation} \label{approach2}
 \psi_{x}[\varphi]
 = e^{S_{\La}[\varphi]} \left( \int [d J] e^{i \varphi \cdot J} \frac{- i J(x) }{J \cdot G \cdot J} \tilde f[J]
 + w_{x}[\varphi] \right),
\end{equation}
where $w_{x}[\varphi]$ is an arbitrary divergenceless vector field according to Eq.~(\ref{decomp2}),
which again can be eventually chosen to improve the behavior of the solution.
Therefore we can see that formally, given an arbitrary (Wilsonian) flow, one can construct a coarse-graining procedure which generates it.
Moreover the coarse-graining procedure is not unique, indeed there are infinitely many. One can take eventually advantage
of this freedom to make the most sensible choice of coarse-graining from the physical point of view.

One can even try to formalize this picture considering the space of Wilsonian actions ${\cal S}$ and the space of
Wilsonian RG flows ${\cal V}$, where each point $p \in {\cal V}$ is the vector field functional of the flow, e.g.,
associated to Eq.~(\ref{eq:generalFlow}).
Then one can consider a fiber bundle space ${\cal F}$ such that, to each point $p$ belonging to its base ${\cal V}$ is
associated a fiber related to the coarse-graining generator $\psi_{x}$.
Like in a gauge theory an infinite set of $\psi_{x}$ is associated to the same vector field $p$
generating the Wilsonian RG flow.
Using dual description one could also associate generalized higher order forms dual to the vector fields and their divergence.

Summarizing, let us remark that this procedure ensures that the action $S_\La$, which satisfies a flow equation which can be casted in the form of Eq.~\eqref{eq:generalCG} thanks to the relation of Eq.~\eqref{approach2}, is related through a functional integral of Eq.~\eqref{partZ} to a partition function independent from $\La$ by construction.
%
%
\section{Proper time flows}

It has been recently proposed~\cite{deAlwis:2017ysy} that another Wilsonian action can satisfy the flow equation
\begin{equation}
 \La \frac{d}{d \La}S_\La[\varphi] = \Tr \ e^{- S^{(2)}_\La[\varphi] /\La^2}.
 \label{eq:PTexp}
\end{equation}
This equation and some generalizations were considered in the past as approximated RG schemes in different kind of contexts. 
It was proven that they could not be interpreted as defining a flow for the IR regulated generating functional of the proper vertices~\cite{Litim:2001hk, Litim:2001ky, Litim:2002hj, Litim:2002xm}.

This flow equation can be formally obtained considering a more general family of PT flows
\begin{equation} \label{eq:genPT_flow}
 \dot{S}_\La[\varphi]
 = \frac{1}{2} \tr \int_{0}^{\infty} \! \frac{ds}{s} \, r_\La(s) \, e^{-s S^{(2)}_{\La}[\varphi]} \, ,
\end{equation}
where $s$ is the Schwinger proper time, the "dot" on the l.h.s. denotes the derivative with respect to the RG "time"
$t=\log{\La/\La_{0}}$ and 
\be
\label{eq:cut}
r_\La(s)=\La \frac{d \reg}{d \La} \,
\ee
with $\reg$ is a suitably normalized regulator function, where $k$ and $\Lambda$ are IR and UV scales respectively.
In particular the flow of Eq.~(\ref{eq:PTexp}) is obtained choosing the regulator function $\reg$ to be
\begin{equation} \label{thetareg}
 \reg = \theta(s - 1/\Lambda^{2}) - \theta(s - 1/k^{2})
\end{equation}
with $k<\Lambda$.

The function $\rho_{k,\Lambda}$ can represent a huge family of regulators.
One can see that making a particular choice already known in the literature it is possible to obtain 
a flow which may look similar to the Wetterich flow for the effective average action. One should not be mislead
by this analogy since our goal is to show that this flow equation can be interpreted only as a Wilsonian action 
(to appear inside a functional integral).

Let us consider the following 1-parameter family of regulators ($m\geq 0$),
\begin{equation}
 \rho_{k, \La}(s ; m) = \frac{\G(m , m s k^{2}) - \G(m , m s \La^{2})}{\G(m)} .
 \label{eq:spectral_rho_gamma}
\end{equation}
Using the properties of the incomplete Gamma function we obtain
\begin{equation} \label{eq:spectral_der_rho_gamma}
 \La \frac{d}{d \La} \rho_{k, \La}(s ; m) = \frac{2}{\G(m)}(m s \La^{2})^{m} e^{- m s \La^{2}} = r_{\La}(s ; m) \, ,
\end{equation}
where we have also put in evidence how this object is actually dependent on the UV cutoff $\La$ only.
We shall denote this PT RG scheme as the A-scheme.

Inserting this choice of regulator in (\ref{eq:genPT_flow}) and performing the Mellin transform we get the following exact RG
equation for the Wilsonian action $S_\La$,
\begin{equation} \label{eq:new-ERGE}
 \La \frac{d}{d \La} S_{\La}[\varphi] = \Tr \left( \frac{m \La^{2}}{S^{(2)}_{\La}[\varphi] + m \La^{2}} \right)^{m} .
\end{equation}
Note that for $m = 1$ this looks like the Wetterich Equation with a massive regulator (which can define a flow for the IR regulated effective average action in low dimensionality).
In the LPA approximation there also exists a specific value of $m = d/2 + 1$ which gives essentially the same flow for the
potential as the Wetterich equation for an optimized regulator.
One should not be misled by these analogies, since $S_\La$
has to be interpreted only as a Wilsonian action 
(to appear inside a functional integral) and not 
as an IR-regulated generator of the 1PI correlators
\cite{Litim:2002hj}.

Note also that for $m \rightarrow \infty$ the scale derivative of $\reg$ becomes a Dirac delta distribution, or equivalently,
one has in~(\ref{eq:new-ERGE}) the representation of an exponential, so the flow equation reduces exactly to the one given
in Eq.~(\ref{eq:PTexp}).

It is also worth to discuss the possible field dependence one could choose to have in the PT regulator. 
This is related to the idea that one can introduce schemes with a different level of spectral adjustement.
Let us start from a generic Wilsonian action $S_{\La}(\varphi)$ casted in the form of a derivative expansion with 
a structure of the Hessian $S^{(2)}_{\La}(\varphi) = Z_{\La}(\varphi) (-\Box) + \cdots$.
The coarse-graining scheme induced by the proper time regulator $\rho(s)$ controls the integration of the UV modes since there is a cutoff related to the condition $s < \frac{1}{\La^{2}}$, which means that modes with $p^{2}$ such that $p^{2} Z_{\La} > \La^{2}$ are roughly integrated out.
Therefore in this scheme the cutoff may be not directly related to the space time scale of the fluctuation, since this scale
is modulated by a flowing field dependent dynamical factor $Z_\La$.
Let us now consider a more spectrally adjusted scheme obtained using a proper time regulator $\rho(Z_{\La} s)$.
In this case the modes are integrated out if $s < \frac{1}{\La^{2} Z_{\La}}$ so that they have momentum $p^{2} > \La^{2}$, 
which is better related to the original idea of cutting off the modes comparing their wavelengths to the sliding scale $1/\La$.
One can observe that it is in the spectrally adjusted scheme that the two steps in the renormalization procedure of (1) coarse-graining and (2) rescaling are tuned to each other as desired for a comparison along the flow.

In the case of such spectrally adjusted proper time regulators there is an ambiguity related to the fact that one can choose to introduce the definition at the level of finite or infinitesimal coarse-graining procedures. 
Inserting the $Z$ dependent spectral adjustement  at the level of Eq.(\ref{eq:spectral_rho_gamma}), which correspond to give a definition at the level of a finite RG flow, one obtains for the incomplete Gamma form the flow equation
\begin{equation} \label{eq:new-ERGEadj}
 \La \frac{d}{d \La} S_{\La}[\varphi]
 = \Tr \left[ \left( 1 + \frac{1}{2} \La \frac{d}{d \La} \log Z_{\La} \right)
 \left( \frac{ m \La^{2} Z_{\La} }{ S^{(2)}_{\La}[\varphi] + m \La^{2} Z_{\La}} \right)^{m} \right].
\end{equation}
In the limit $m \to \infty$ one gets
\begin{equation} \label{eq:expadj}
 \La \frac{d}{d \La} S_{\La}[\varphi]
 = \Tr \left[ \left( 1 + \frac{1}{2} \La \frac{d}{d \La} \log Z_{\La} \right)
 \ e^{- \frac{S^{(2)}_{\La}[\varphi] }{ \La^{2} Z_{\La}} } \right].
\end{equation}
We shall denote this class of PT RG schemes as the B-scheme.

A parallel reasoning carried on at the level of Eq.~(\ref{eq:spectral_der_rho_gamma}), correponding to an infinitesimal RG flow,  leads instead
to similar relations, but without the term containing the scale derivative of the logarithm $\La \frac{d}{d \La} \log Z_{\La}$.
The latter case (which we denote C-scheme) has been the scheme, which after a further approximation as explained in~\cite{BLPV} (simplified C-scheme), 
has been mostly used in the PT literature~\cite{Litim:2010tt, Mazza:2001bp, Bonanno:2000yp}.

Going back to the goal of this section, the path to interprete in the Wilsonian sense the general proper time flow Eq.~(\ref{eq:genPT_flow}) discussed above,
consists in rewriting it in the form given in the Eqs.~(\ref{eq:generalFlow})
or~(\ref{eq:generalCG}).
Specific realizations can be  further investigated  looking for a solution of the following equation
\begin{equation} \label{eq:PTcond}
 f[\varphi]
 = - e^{- S_{\La}[\varphi]} \frac{1}{2} \tr \int_{0}^{\infty} \frac{d s}{s}
 \left[ r_{\La}(s) e^{-s S^{(2)}_{\La}[\varphi]} \right] \, .
\end{equation}

Going back to the goal of this section, in order to interpret in the Wilsonian sense the general PT flow Eq.~(\ref{eq:genPT_flow}),
we would have to rewrite it in the form given in the Eqs.~(\ref{eq:generalFlow})
or~(\ref{eq:generalCG}).
Thus we must look for a solution of the following functional equation
\begin{equation} \label{eq:condpsi2}
 \int \! d x \, \frac{\delta}{\delta \varphi(x)} \left( \psi^\La_{x}[\varphi] e^{- S_{\La}[\varphi]} \right)
 = - e^{- S_{\La}[\varphi]} \frac{1}{2} \tr \int_{0}^{\infty} \frac{d s}{s}
 \left[ r_{\La}(s) e^{-s S^{(2)}_{\La}[\varphi]} \right] \, .
\end{equation}
Before moving to this task let us make a comment. The existence of a solution would make it possible to interpret this particular PT regulated action $S_\La$, as a Wilsonian action, i.e. an action which, inserted in a functional integral,
not only generates the partition function $Z$, but also all the possible correlators (connected and not connected) with momenta below the scale $\La$, i.e.
\begin{equation}
\langle O_1(x_1) \cdots O_n(x_n) \rangle=\frac{1}{Z} \int [d \varphi]_\Lambda e^{-S_\La[\varphi]} O_1(x_1) \cdots O_n(x_n) \,.
\end{equation}
From the structure of the flow equation one notes that in general, contrary to the Polchinski action, this action gets along the flow contributions which results into 1PI non local vertices. 
We stress that the relation between the PT regulated $S_\La$ and the effective action $\Gamma$ (eventually IR regulated) 
is not trivial and certainly not so simple as for the Polchinski Wilsonian action.

\subsection{The free theory case}
%
In the previous section we have seen that a formal solution for the field redefinition can be mathematically constructed, independently from the requirements
to interprete it as a good Wilsonian coarse graining procedure from a physical point of view. 
Clearly one would like to avoid possible pathological definitions.
This is a general question not only related to the specific case of a proper time flow.
Tipically one would require a quasi local field redefinition, which can be
written in power series of the field in a derivative expansion. On the other hand this property is not expected to be physically necessary
in presence of new degrees of freedoms, such as bound states, in the spectrum of the theory. 
In this case  some non local structures are expected to appear along the RG flow in the IR regime.
A first check consists to see that quasi locality is at least maintained in a free theory, which is not a proper-time flow specific property. 

Let us then investigate in detail the case of a free quadratic action for the case of a generic cutoff. 
Using a natural compact notation we have
\begin{equation}
 S[\varphi] = \frac{1}{2} \varphi \cdot (-\Box)_{\Lambda} \cdot \varphi , \quad
 f[\varphi] = - \frac{1}{2} e^{-S_{\La}[\varphi]} \, {\rm Tr} \int_{0}^{\infty} \frac{d s}{s}
 \left[ r_{\La}(s) e^{-s (-\Box)_{\La}} \right]
\end{equation}
so that one can write
\begin{equation}
 \tilde f[J]
 = - \frac{1}{2} {\rm Tr} \int_{0}^{\infty} \frac{d s}{s}
 \left[ r_{\La}(s) e^{- s (- \Box)_{\La}} \right] \left( {\rm Det} G\right)^{\frac{1}{2}} e^{- \frac{1}{2} J \cdot G \cdot J} \, ,
\end{equation}
which represents the source of the Poisson-like equation~(\ref{infpoisson2}). Its solution gives, for the "potential" $h$,
\bea \label{sol2}
 h[\varphi]
 &=& \frac{1}{2} {\rm Tr} \! \int_{0}^{\infty} \frac{d s}{s} \!
 \left[ r_{\La}(s) e^{- s (- \Box)_{\La}} \right]
 \int [d J] e^{i \varphi \cdot J} \frac{\left( {\rm Det} G\right)^{\frac{1}{2}}}{J \cdot G \cdot J} e^{- \frac{1}{2} J \cdot G \cdot J} \nn
&=& {\cal N} \!\! \int [d \tilde{J}] e^{i  \tilde{J} \cdot G^{-\frac{1}{2}} \varphi}
 \frac{1}{ \tilde{J} \cdot \tilde J} e^{- \frac{1}{2} \tilde{J} \cdot \tilde{J}} ,
\eea
where $\tilde{J} = G^{\frac{1}{2}} J$ and we have defined the normalization factor
\begin{equation}
 {\cal N}
 = \frac{1}{2} {\rm Tr} \int_{0}^{\infty} \frac{d s}{s} \left[ r_{\La}(s) e^{- s (- \Box)_{\La}} \right]\,.
\end{equation}
The next step is to take a functional derivative to construct $v_{x}$ and the infinitesimal Wilsonian
field redefinition $\Psi_{x}$ as well. This means that in the integral in Eq.~(\ref{sol2}) the singular region in the origin is harmless and
one could replace $e^{i \tilde{J} \cdot G^{- \frac{1}{2}} \varphi} \to \left( e^{i \tilde{J} \cdot G^{-\frac{1}{2}} \varphi }
- 1 - i \tilde{J} \cdot G^{- \frac{1}{2}} \varphi \right)$, given that only $\tilde{J}$ even terms in the integrand give non zero contributions.

To compute the integral it is simpler to introduce a parameter $a$, 
which will be set to $\frac{1}{2}$ at the end in the previous formal solution for $h[\varphi]$:
\begin{equation} \label{sol2a}
 I_{a}[\varphi]
 = {\cal N} \int [d \tilde{J}] e^{i \tilde{J} \cdot G^{-\frac{1}{2}} \varphi}
 \frac{1}{\tilde{J} \cdot \tilde{J}} e^{- a \tilde{J} \cdot \tilde{J}} \,.
\end{equation}
Taking a derivative w.r.t.~$a$ and performing the $\tilde{J}$ functional integration one has
\begin{equation} \label{sol2bis}
 - \frac{d}{d a} I_{a}[\varphi]
 = {\cal N} \int [d \tilde{J}] e^{i \tilde{J} \cdot G^{- \frac{1}{2}} \varphi} e^{- a \tilde{J} \cdot \tilde{J}}
 = {\cal N} \frac{\sqrt{\pi}}{\sqrt{a}} e^{- \frac{ \varphi (- \Box)_{\La} \varphi}{4 a}} .
\end{equation}
Therefore $I_a$ can be thought of as a function of $S_\Lambda[\varphi]=\frac{1}{2}\varphi \cdot (-\Box)_{\La} \cdot \varphi$.
Re-integrating back in $a$ one finds
\begin{equation}
 I_{a}[\varphi]
 = - {\cal N} \left[ 2 e^{- \frac{\varphi (- \Box)_{\La} \varphi}{4 a} } \sqrt{\pi a}
 + \pi \sqrt{\varphi (- \Box)_{\La} \varphi} \,\, {\rm erf} \left( \sqrt{ \frac{\varphi (-\Box)_{\La} \varphi}{4 a} } \right)
 + c[\varphi] \right] \, ,
\end{equation}
where the last term is an $a$-independent functional. From a direct integration it results to be a simple constant
and can be discarded.
Setting $a = 1/2$ and ignoring also $w_{x}[\varphi]$ in Eq.~(\ref{decomp2}) one finally arrives at
\begin{eqnarray}
 \!\!\!\! \psi_{x}[\varphi]
 &=& e^{-\frac{1}{2} \varphi (-\Box)_{\La} \varphi}  G_{x y} \frac{\delta h[\varphi]}{\delta \varphi(y)}
 = - {\cal N} \pi \frac{e^{-\frac{1}{2} \varphi (-\Box)_{\La} \varphi}}{ \sqrt{ \varphi (-\Box)_{\La} \varphi } }
 {\rm erf} \left( \sqrt{ \frac{\varphi (- \Box)_{\La} \varphi}{2} } \right) \varphi_{x}
 \\
&=& - {\cal N} \sqrt{2 \pi} \left( 1 - \frac{2}{3} \varphi (- \Box)_{\La} \varphi
 + \frac{7}{30} ( \varphi (- \Box)_{\La} \varphi )^{2} - \frac{2}{35} ( \varphi (- \Box)_{\La} \varphi)^{3} + \cdots
 \right) \varphi_{x} \, . 
 \nonumber
\end{eqnarray}
We note that in this free theory the coarse-graning is, as said, a function of the free Wilsonian action, which is non local, however this non locality is harmless.

Going to an interacting case any analysis becomes extremely complicated and we shall not attemp to make further analysis.
As said not always we should expect quasi locality, e.g. in strongly interacting theories where bound states appear in the
spectrum, for which the flow equation is also bound to be non-local.%
Given that infinitely many coarse-graining schemes can give the same flow because of
the freedom to introduce the divergenceless vector in field space,
as defined in Eq.~(\ref{decomp2}), one may eliminate some pathological non-localities.
Similar considerations can be applied in a quasi local regime, i.e. when the action can be written as a power series in derivatives.
\section{Discussion}
We have discussed how to construct in principle a solution of the inverse RG problem consisting on finding the coarse-graining given a Wilsonian RG flow.
This procedure can be applied to a family of known proper-time regulated RG flows, which then could represent the non perturbative flow of a Wilsonian action.

\begin{figure}[htpb]
\begin{center}
\vskip -2cm
\includegraphics[width=8cm]{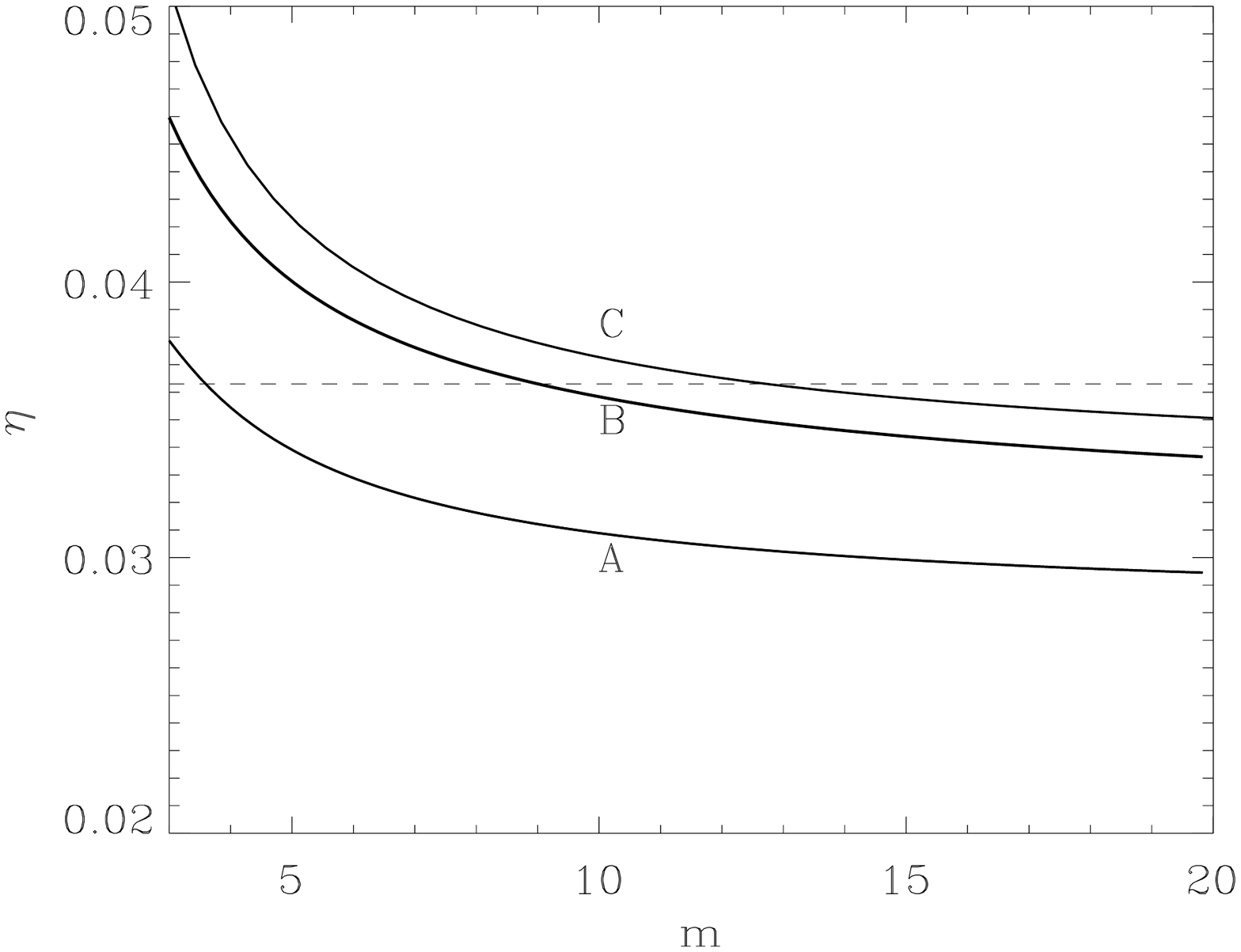}
\vskip -3cm
\caption{The anomalous dimension $\eta$ for the 3D Ising Universality class obtained in the three different schemes A,B and simplified-C. The dashed line correspond to the best known analytical and montecarlo results. }
\label{critical}
\end{center}
\end{figure}

Considering a one parameter ($m$) family of proper-time flows, one can further specify different kind of spectral adjustements. 
We denoted the case with no further adjustement as the A-scheme, and introduced two other schemes, B and simplified-C (this one well investigated in the literature), 
such that the proper-time regulator gives a coarse-graining in the momenta according to the sliding UV scale $\La$.
We have applied these flows to the study of the Ising Universality class in three dimensions~\cite{BLPV}. 
As an example we report in Fig.~\ref{critical} the anomalous dimension (exponent $\eta$) for the three schemes. 

We expect a much milder $m$-dependence in the results for the critical exponent $\nu$ in the three different schemes. Such analysis will be presented elsewhere.

Further theoretical investigation on the field redefinitions to be associated to the coarse-graining schemes are needed to better clarify general properties of Wilsonian flows. 
Moreover several features of PT Wilsonian flows should be further tested, through the study of different critical systems.
\section*{Acknowledgments}

The author is grateful to the organizers of the XXVth edition of the conference in Vietri "PAFT 2019, Current problems  in theoretical physics",
who were also providing a nice atmosphere for discussions.

%
%
%


\end{document}